\documentclass{WileyMSP-template}

\usepackage{soul,amsmath,color}

\usepackage{braket}
 \usepackage[english]{babel}

\begin{document}

\pagestyle{fancy}
\rhead{
}

\title{Self-referenced subcycle metrology of quantum fields}

\maketitle



\author{Sinan G\"{u}ndo\u{g}du*}
\author{St\'{e}phane Virally}
\author{Marco Scaglia}
\author{Denis V. Seletskiy*}
\author{Andrey S. Moskalenko*}



%
%

\begin{affiliations}
Sinan Gundogdu, Andrey S. Moskalenko\\
Department of Physics, KAIST, Daejeon 34141, Republic of Korea\\
Email Address: moskalenko@kaist.ac.kr

Sinan Gundogdu\\
Center for Theoretical Physics of Complex Systems, Institute for Basic Science (IBS), Daejeon 34126, Republic of Korea\\
Email Address: gu.sinan@gmail.com

St\'{e}phane Virally, Marco Scaglia, Denis V. Seletskiy\\
femtoQ Lab, Department of Engineering Physics, Polytechnique Montr\'{e}al, Montr\'{e}al, QC H3T 1J4, Canada\\
Email Address: denis.seletskiy@polymtl.ca

\end{affiliations}


\begin{abstract}

We propose and analyze a new time-domain method for subcycle metrology of quantum electric fields using a combination of a 3rd order nonlinear optical process and homodyne detection with a local oscillator (LO) field. The new method enables isolation of intrinsically weak quantum noise contribution by subtraction of the shot noise of the LO on a pulse-by-pulse basis. Together with the centro-symmetric character of the nonlinearity, our method unlocks novel opportunities toward terahertz and mid-infrared quantum field metrologies.

\end{abstract}


\section{Introduction}

Homodyne detection (HD)~\cite{Collett1987} is a central technique of signal analysis in quantum optics. A quantum field under study interferes on a photo-diode with a strong classical mode $E_\mathrm{LO}$, termed local oscillator (LO)~\cite{Lvovsky2009}. Fitting for the visible and near-infrared frequency bands, HD of quantum fields in the range from mid-infrared (MIR) to terahertz (THz) is challenged by a general unavailability of bright sources of classical LO and high-efficiency photo-diodes. Despite such challenges, interest in these frequency bands is largely motivated by direct sensing of unique signature absorption features in solid-state, liquid and gaseous targets~\cite{Schliesser2012}. Parallel developments in short-pulse lasers ~\cite{Krausz2009} and measurement techniques, such as electro-optic sampling (EOS) \cite{Wu1995,Gallot1999,Leitenstorfer1999}, have stimulated metrology of classical fields at terahertz (THz) to mid-infrared (MIR) frequencies, and above ~\cite{Wu1997,Kubler2004, Gaal2007, Sell2008, Keiber2016, Riek2017EJP}. As an EOS setup, consider a signal $E_T$ at a low center frequency $\Omega$ (e.g. THz or MIR) and a probe pulse $E_P$ at a high center frequency $\omega_0$ (e.g. near-infrared, NIR) and bandwidth $\Delta\omega$ that co-propagate inside of a second-order ($\chi^{(2)}$) nonlinear crystal with electro-optic activity. A mixing product of this interaction, $E_T\cdot E_P$, can span over frequencies $\omega_0 \pm \Omega$, in spectral overlap with the probe pulse when $2\Omega < \Delta\omega$. The resulting interference is exploited for the EOS detection of instantaneous amplitude $E_T$, resolved as a function of time difference $\tau$ between $E_P$ and $E_T${, provided that the bandwidth of the probe field $E_P$ is larger than $2/T$, where $T$ is the characteristic cycle period of the signal field $E_T$}. In contrast to HD {and its generalizations to the pulsed LO case~\cite{Zavatta2004,Hansen2001,Eckstein2011,Anderson1997,Wasilewski2006,Raymer2020}}, EOS is an intrinsically time-domain technique, alleviating the need for direct detection of the signal field.\par

Recent demonstrations ported EOS to the quantum regime, showing direct sampling of a quantum vacuum field ~\cite{Riek2015,Moskalenko2015}, and even measuring the spatio-temporal correlations~\cite{Chelmus2019} and causal structure of the electromagnetic ground state~\cite{Settembrini2022}. Such developments promise direct routes toward MIR and THz quantum sensing technologies, while the time-domain character motivates new metrology protocols~\cite{Virally2019,Sulzer2020,Virally2021} as well as a path toward experimental quantum electrodynamics in space-time~\cite{Riek2017,Kizmann2019,Lindel2020,Kizmann2022,Settembrini2022}. Despite this progress, time domain quantum photonics faces a few outstanding challenges. On the one hand, strong phonon-polariton dispersion in the $\chi^{(2)}$ crystal used in EOS makes the detection of signals around the Reststrahlen band frequencies challenging~\cite{Gallot1999,Leitenstorfer1999}. On the other, the fractional content of the quantum contribution to the variance of the total detected signal amounts to only a few percent of the shot-noise of $E_P$~\cite{Moskalenko2015}. Experimental differentiation between the two contributions must be exquisitely precise~\cite{Riek2015}, ideally requiring pulse-by-pulse comparison~\cite{Riek2017} of events with and without the quantum contribution, a feat that is not currently available experimentally. \par

In this Letter, we propose a new scheme for time-domain metrology of quantum signals based on the third-order ($\chi^{(3)}$) nonlinear interaction between quantum $E_T$ and classical $E_P$, admitting full access to the term carrying interference of $E_\mathrm{LO}$ and the signal fields{~\cite{Li2015,Karpowicz2008,Tomasino2021}}. In the four-wave mixing, the THz-induced second-harmonic (TFISH) signal arises from the nonlinear mixing product $E_P \cdot E_P \cdot E_T$ at frequencies $2\omega_0 \pm \Omega$, which can be superposed in a background-free manner with the $E_\mathrm{LO}$ centered at $2\omega_0$. We show that this freedom opens an elegant opportunity for a direct self-referenced measurement of the quantum contribution to the signal variance, based on a shot-by-shot carrier-envelope phase (CEP) modulation on $E_\mathrm{LO}$ of a free-running frequency comb~\cite{Fehrenbacher2015}. Furthermore, dipole inactivity of optical phonons in common inversion-symmetric materials positions the $\chi^{(3)}$-based scheme for efficient field-resolved detection in the 5-15 THz band, which is generally problematic for a host of EOS detection crystals~\cite{Leitenstorfer1999}. Finally, the new scheme does not require analysis of the polarization state of the probe, relaxing constraints on broadband polarization optics~\cite{Sulzer2020}.
\begin{figure}[t]
    \centering
    \includegraphics[width=0.9\columnwidth]{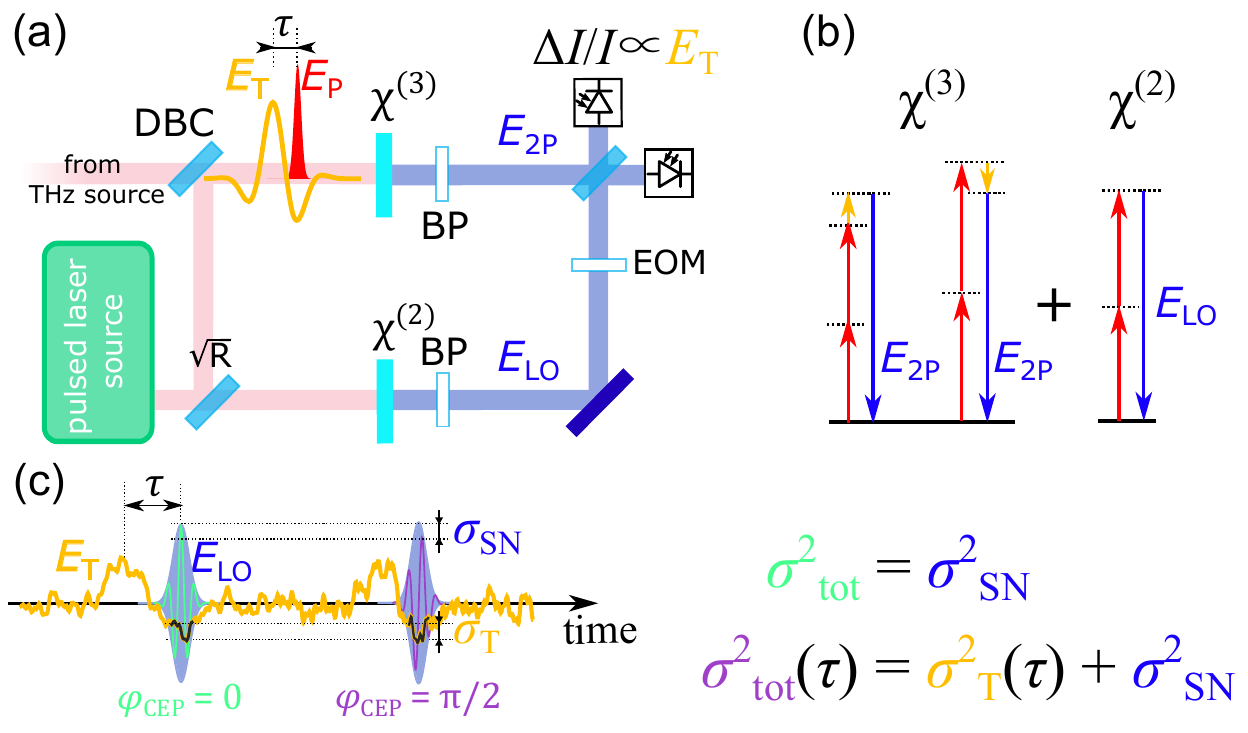}
    \caption{ (a) Proposed metrology scheme. The output of a near-infrared (NIR) pulsed laser source is split in two arms. Upper arm: $E_T$ [THz signal (orange)] and a part of $E_P$ [NIR probe (red)] are time-delayed by $\tau$, overlapped via a dichroic beam combiner (DBC), and mixed in a $\chi^{(3)}$ crystal to generate $E_{2P}$ signal (blue). Lower arm: $E_{\rm LO}$ [Local oscillator (LO, blue)] is derived from a frequency-doubled remainder of $E_P$ and passes through an electro-optic modulator (EOM). After bandpass filters (BP), $E_{\rm LO}$ and $E_{2P}$ are mixed in a balanced homodyne detector, schematized as a superposition of $\chi^{(3)}$ and $\chi^{(2)}$ processes in (b). (c) EOM-shifted $E_{\rm LO}$ contains a $\pi/2$ flip of the CEP between consecutive pulses, so that their total noise variance $\sigma_\mathrm{tot}$ is oscillating between having only shot-noise contribution $\sigma_{\rm SN}$ or also containing the variance of the quantum field $\sigma_T(\tau)$.}
    \label{fig:setup}
\end{figure}

\section{Generation of the TFISH and LO fields}
\textbf{Figure \ref{fig:setup}} shows the schematic of the proposed metrology scheme. It consists of an interferometer with two paths. The upper path combines $E_{T}$, the THz signal to be characterized, and one part of the NIR probe pulse $E'_P=\sqrt{R}E_P$, both passing through a $\chi^{(3)}$ nonlinear crystal to generate the TFISH signal $E_{2P}$. In the lower path, the second part of the probe $\sqrt{1-R}E_P$ undergoes a broadband second harmonic generation (SHG) process in a $\chi^{(2)}$ nonlinear crystal, to generate the LO pulse $E_{\rm LO}$ centered at a frequency $2\omega_0$.  $E_{2P}$ is mixed with $E_{\rm LO}$ at a beamsplitter, and analyzed by the balanced homodyne detection technique.
The measurement system
is based on
a free running frequency comb, whereas an external EOM provides an opportunity to control the carrier-envelope phase shift $\Delta \phi_{\rm{CEP}}$ of the LO field on a pulse-by-pulse basis. This feature is exploited for the self-referenced detection, where the difference in the variance sampled by two adjacent LO pulses isolates the quantum contribution (\textbf{Figure \ref{fig:setup}}c), as detailed below.

The total electric field passing through the $\chi^{(3)}$ crystal, $E$, can be represented as a sum:
$E= E'_{P} + E_{2P}+ E_T.$
When $E_T$ corresponds to a quantum signal, its properties together with those of $E_{2P}$ and $E$, should be described in terms of operators denoted as $\hat{E}_T$, $\hat{E}_{2P}$ and $\hat{E}$, respectively. Since the probe is a strong coherent field, it is sufficient here to describe it classically using the corresponding mean value $\langle \hat{E_P'}\rangle\equiv E_P'$. Thus, the generation of $\hat{E}_{2P}$ is described by the third order nonlinear polarization:
\begin{equation}\label{Eq:P_NL_chi3}
\hat{P}^\mathrm{NL}_{2P}= 3\epsilon_0 \chi^{(3)}R
 E_{P}^{\prime (+)} E_{P}^{\prime(+)} \big[\hat E_{T}^{(+)} +\hat E_T^{(-)}\big]+\mathrm{H.c.},
\end{equation}
where we have decomposed $E'_P$ ($\hat{E}_T$) into its positive $E_P^{\prime(+)}$ ($\hat E_T^{(+)}$) and negative $E_P^{\prime (-)}$ ($\hat E_T^{(-)})$ frequency parts \cite{Glauber1963}. $\epsilon_0$ is the vacuum permittivity.  We have assumed that the interaction can be well described by an effective $\chi^{(3)}$ constant, neglecting its frequency dependence. Further, we consider such a crystal orientation that it is sufficient to include only one linear polarization component for each of the involved fields. We also ignore any other third order nonlinear polarization terms since they are comparably much weaker, due to the fact that $E'_P$ is a strong coherent field and $\hat{E}_T$ is a weak quantum signal, or the corresponding generated field contributions are removed by an appropriate bandpass filter before the final beamsplitter. Under these general assumptions, we can describe the electromagnetic wave propagation in the $\chi^{(3)}$ crystal using the inhomogeneous wave equation within the slowly varying amplitude approximation (SVAA) for plane waves. We decompose all fields, $X =E_P',\hat E_T, \hat E_{2P}, \hat{P}_\mathrm{NL}$, into forward-propagating plane waves as
$X(z,t)=\int_{0}^\infty X(z,\omega)e^{i [ k(\omega) z-\omega t]}d\omega+\mathrm{H.c.}\;$
using a convention $X(z,\omega)\equiv0$ for $\omega<0$, $X^{(+)}(z,\omega)\equiv X(z,\omega)$, and $X^{(-)}(z,-\omega)=\big[X(z,\omega)^{(+)}\big]^\dagger$, which has to be taken into account for convolutions we define below. Then solving the propagation equation for the TFISH under the assumptions of small crystal thickness $d$ and negligible depletion of the probe, we obtain (cf. Supporting Information):
\begin{equation}\label{Eq:E2P_result}
 \hat{E}_{2P}(\omega_2) =
i d A_P^2 C(\omega_2)\big (\mathcal{R}* (\hat E^{(+)}_{T}+ \hat E^{(-)}_{T})\big)(\omega_2) +\hat{E}_{B}(\omega_2).
\end{equation}
Here $*$ denotes convolution and $ C(\omega)=3 \chi^{(3) }R\omega/\big[2 c n(\omega)\big]$, where $n(\omega)$ is the frequency-dependent refractive index and $c$ is the speed of light in vacuum.
We decomposed $E_P=A_P f(\omega)$ into its (real) amplitude $A_P$ 
and (generally complex) normalized frequency distribution $f(\omega)$,
defining a gating function $\mathcal{R}(\omega_2)=(f * f )(\omega_2)e^{-i\omega_2\tau}$. The time $\tau$ corresponds to the center of the probe pulse relative to the incoming THz field (\textbf{Figure \ref{fig:setup}}a).
Finally,
$ \hat{E}_{B}(\omega_2)$ is the co-propagating background vacuum field, existing even in the absence of the probe field. As we will see below, this contribution can be directly characterized in the current setup by studying
{the signal variance} when
{$\phi_\mathrm{CEP}=0$}
(cf. \textbf{Figure \ref{fig:setup}}c) is induced on the LO field.  
In the above derivation, we have neglected the back-action effect of the $\chi^{(3)}$ interaction on the co-propagating THz field $\hat{E}_{T}$ \cite{Thiago_BA}. That is appropriate if the first term on the right-hand side of Equation \eqref{Eq:E2P_result} represents a small correction to the second term {and is in accordance with our evaluation of the variance signals}.

\section{Homodyne detection}\label{Sec:Homodyne}
As described, homodyne detection is enabled by letting $\hat{E}_{2P}$ interfere with the LO field $E_{\rm LO}$ which is produced via the SHG in the $\chi^{(2)}$ crystal. The SHG process can be realized with a high conversion efficiency $\eta_2$, and the resulting mode $f_{\rm LO}(\omega)$ of the LO still closely resembles the mode of the TFISH field. For the details of a possible realization, see Supporting Information. Since the LO represents a strong coherent field, we can describe it classically for the homodyning part of the setup. At the beamsplitter before the detectors, we can express it in the frequency domain as
$E_{\rm LO}(\omega_2)
=A  e^{-i \phi} e^{-i \omega_2\tau}f_{\rm LO}(\omega_2)$,
where $A=\sqrt \eta_2 \sqrt{1-R}A_P$ is the amplitude and $\phi$ is a variable CEP induced by the EOM. An elegant alternative realization of such phase shift between the two arms of the setup can be provided by a dual frequency comb \cite{Coddington2016}.
We assume the LO has effectively the same time delay $\tau$ as the TFISH field because of the equal optical path lengths corresponding to the upper and lower arms of the setup. To assure that this is completely fulfilled in experiment, an additional tuning can be introduced by inserting an auxiliary delay line. 

The operator for the signal we are interested in is given by the difference of number of photons between the pair of balanced photodetectors, $\hat{N}_2-\hat{N}_1\equiv\hat{S}_{\rm hom}$, with
\begin{equation}\label{Eq:S_hom1}
\hat{S}_{\rm hom}\!=C'\!  A\!\!  \int_{0}^{\infty}\!\frac{d\omega_2}{\hbar \omega_2}n(\omega_2)\eta(\omega_2)e^{i \phi} f_{\rm LO}^{*}(\omega_2) \hat{E}_{2P}(\omega_2)+\mathrm{H.c.},
\end{equation}
where $C'=4 \pi c  \epsilon_0 F$, $F$ is the effective detection area and $\eta(\omega)$ is the frequency-dependent quantum efficiency of the photodetectors.
Using the decomposition  of $\hat{E}_{2P}$ given by Equation \eqref{Eq:E2P_result}, we can write $\hat{S}_{\rm hom}$ as $\hat{S}_{\rm hom}=\hat{S}+\hat{S}_B$, with $\hat{S}_B$ corresponding to the background vacuum signal, and
\begin{equation}\label{Eq:S_final}
\hat{S}(\tau)=
 C'' \int_{0}^{\infty}
 \!\!d\Omega\;
  \mathcal{G}_\phi(\Omega)e^{-i\Omega \tau}\hat{E}_T(\Omega)+\mathrm{H.c.},
\end{equation}
where $C''=6 \pi \epsilon_0 F d \chi^{(3)}\sqrt{\eta_2}\,R\sqrt{1-R}\,A_P^3/\hbar $ and the resulting gating function, limiting the effective integration range in Equation \eqref{Eq:S_final} to THz frequencies, is given by
\begin{equation} \label{Eq:G}
\begin{split}
  &\mathcal G_\phi(\Omega)=ie^{i \phi}
\mathcal{G}_-(\Omega)
- ie^{-i \phi }
\mathcal{G}_+^*(\Omega)\;;\\
&\mathcal{G}_\pm(\Omega)=\int_{0}^{\omega_\mathrm{cut}}\!\!d\omega_2\;
\eta(\omega_2)f_{\rm LO}^{*}(\omega_2) (f*f)(\omega_2\pm\Omega)\;.
\end{split}
\end{equation}
We have introduced $\omega_\mathrm{cut}$ to represent an optional upper spectral limit for the collected NIR photons, as discussed below.
Deriving Equation \eqref{Eq:S_final}, we switched the order of integrations between the integral of Equation \eqref{Eq:S_hom1} and that of the convolution coming there from Equation \eqref{Eq:E2P_result}.
Casting the electric field operator in terms of creation $a^\dagger_T(\Omega)$ and annihilation $a_T(\Omega)$ operators, leading to  $\hat E_T(\Omega)=-i \sqrt{\hbar \Omega/C' n(\Omega)}\,\hat a_T(\Omega)$ for $\Omega>0$,
the TFISH induced contribution to the homodyne signal can be rearranged as
  \begin{align}\label{Eq:S_a}
 \begin{split}
 \hat{S}(\tau)=
\frac{-i C''}{\sqrt{C'}} \int_{0}^{\infty}
\frac{d\Omega \sqrt{\hbar\Omega}}{\sqrt{ n(\Omega)} }
\mathcal{G}_\phi(\Omega)e^{-i\Omega \tau} \hat{a}_T(\Omega)
     + \mathrm{H.c.}\;.
 \end{split}
 \end{align}
In order to obtain the quantum statistics of the operator $\hat S(\tau)$ at each time delay $\tau$ from the general Equation \eqref{Eq:S_a} or \eqref{Eq:S_final} and thus to get access to the time-resolved properties of the quantum field $\hat{E}_T$, we need to assume a particular form of this field.

The simplest case for the consideration is provided by the bare THz vacuum field. In this case, the
mean values of both contributions to $\hat{S}_\mathrm{hom}$ vanish. Further, since these contributions are determined by creation/annihilation operators stemming from different frequency ranges, they are uncorrelated. Thus, the total variance is given by the sum of the variance of the TFISH part,
\begin{equation}\begin{split}
 &\langle \hat{S}^2\rangle_{\rm vac}\equiv \sigma_T^2
 = \langle 0|\hat{S}^2 |0\rangle=\frac{C''^2}{C'}
\int_{0}^{\infty}\frac{d\Omega\,\hbar\Omega }{n(\Omega)}
\left|  \mathcal G_\phi(\Omega)\right|^2,
\end{split} \end{equation}
and of the variance originating from  the background vacuum in the range of $\omega_2$ frequencies (here we assume $\omega_\mathrm{cut}=\infty$),
\begin{equation}\label{Eq:background_variance}
\langle \hat{S}_B^2\rangle\equiv\sigma_\mathrm{SN}^2=   A^2 C'\int_0^\infty\frac{d\omega_2}{\hbar \omega_2}|f_{\rm LO}(\omega_2)|^2 \eta^2(\omega_2)n(\omega_2).
 \end{equation}
Both variances are independent of the time delay $\tau$.
The TFISH part is determined by the properties of the gating function $\mathcal G_\phi(\Omega)$, which follow from the relation between $\mathcal G_+(\Omega)$ and $\mathcal G_-(\Omega)$ and can be influenced by the phase shift $\phi$. When the temporal profiles of the TFISH and SHG signals coincide, we have $\mathcal G_+(\Omega)=\mathcal G_-^*(\Omega)$. Then $\mathcal G_\phi(\Omega)$ is real and vanishes for $\phi=0$. Its absolute value is maximized for $\phi=\pm\pi/2$, with $|\mathcal G_\phi(\Omega)|=|\mathcal G_+(\Omega)+\mathcal G_-^*(\Omega)|=2|\mathrm{Re}(\mathcal G_+(\Omega))|$, representing the optimal configuration for the sampling of the THz vacuum.  {With $\hat{S}_\mathrm{hom}=\hat{S}_B$ for $\phi=0$, the corresponding measurement outcomes can be used for the elimination of the background NIR vacuum contribution on the pulse-by-pulse basis (self-referencing), alternating the
CEP
by $\pi/2$ between the pulses (see \textbf{Figure \ref{fig:setup}}c).}

In order to sample THz quantum fields beyond the bare vacuum, we need to get access to both generalized quadratures of the sampled field in the time domain \cite{Kizmann2019,Virally2019,Sulzer2020}. This is possible to achieve with a variation of the proposed setup, introducing an
asymmetry between $\mathcal G_+(\Omega)$ and $\mathcal G_-(\Omega)$ contributions to $\mathcal G_\phi(\Omega)$. One of the easiest ways to realize this, suitable for our discussion 
here, is based on the cuts in the spectra of the detected photons that can be implemented via the corresponding frequency bandpass filters \cite{Sulzer2020,Kizmann2022}. Looking at the second line of Equation \eqref{Eq:G} one can anticipate that, e.g., for a bandpass filter cutting the frequencies above the central frequency of the LO, $\mathcal G_-(\Omega)$ dominates over $\mathcal G_+(\Omega)$ in terms of the absolute magnitude and $\mathcal G_\phi(\Omega)$ becomes complex. In the easiest case, it can be written as $\mathcal G_\phi(\Omega)=|\mathcal G_\phi(\Omega)| e^{i\theta}$, where the phase $\theta=\theta(\phi)$ is uniquely determined by the phase $\phi$ (and vice versa) whereas being independent of frequency. In particular, we have $\theta=\pi/2$ for $\phi=0$ and $\theta=0$ for $\phi=-\pi/2$. Then we can introduce operators
$\hat S_0(\tau)$ and $\hat S_{\pi/2}(\tau)$, with $\phi=-\pi/2$ and $\phi=0$ in Equation \eqref{Eq:S_final}, respectively. These operators
determine both generalized quadratures of the sampled quantum field, up to
normalization prefactors.
 \begin{figure}[t!]
    \centering
    \includegraphics[width=0.9\columnwidth]{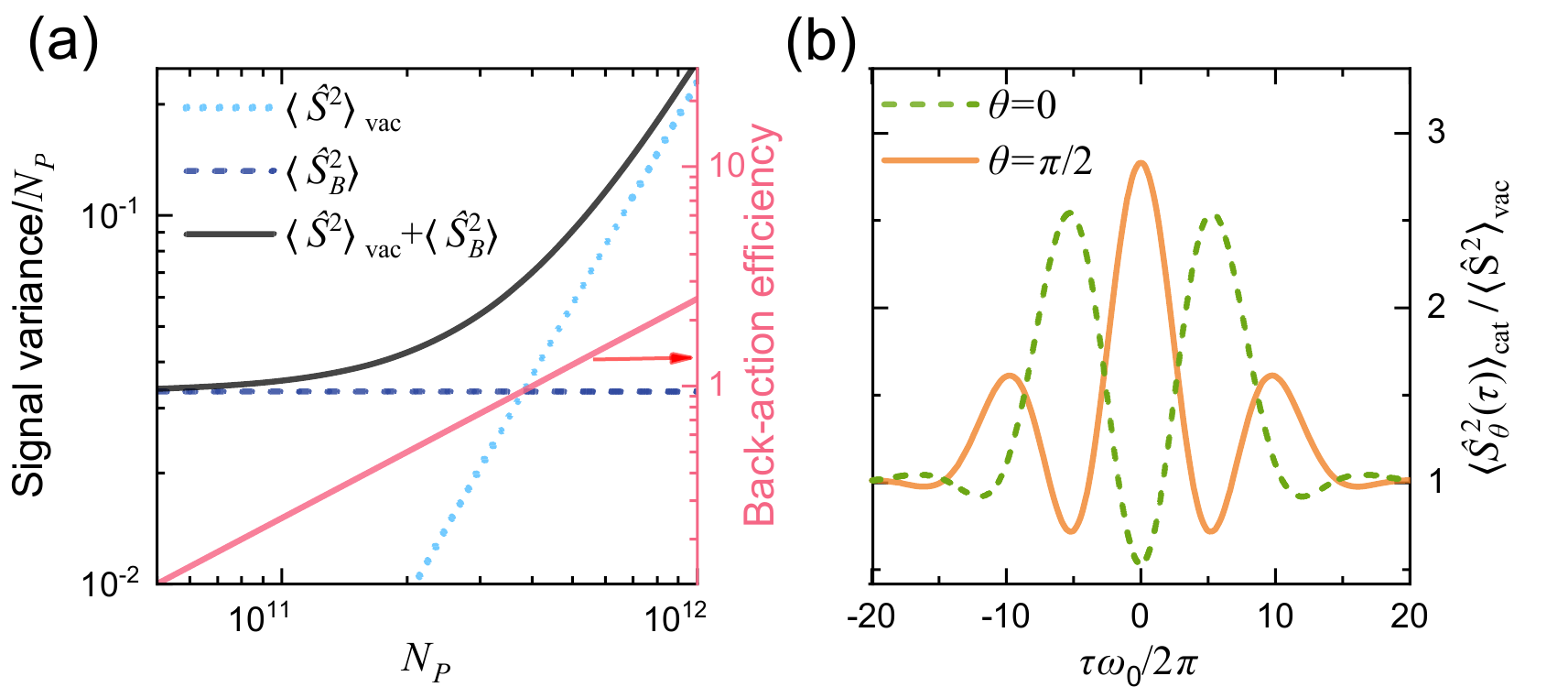}
    \caption{(a) Normalized (by the mean probe photon number $N_P$) signal variance arising from the THz vacuum $\langle \hat S^2 \rangle_{\rm vac}$ and the background LO vacuum $\langle \hat S_{B}^2\rangle$ as well as the corresponding (normalized) total variance are shown as functions of $N_P$, along with a perturbative estimation for the back-action efficiency of the $\chi^{(3)}$ process. Parameter values: $\eta$=1, $\eta_2$=0.1, $R$=2/3 , $d=12~\mu$m, $\sigma/(2\pi) =31$~THz, $\omega_0/(2\pi)=193$~THz, $\chi^{(3)}=2.5 \times 10^{-21}$~m$^2$/V$^2$, $n$=2.4, $F=9~\mu$m$^2$.  (b) TFISH signal variance for a broadband cat state, {$\langle \hat S_{\theta}^2(\tau)\rangle-\langle \hat S_{\theta}(\tau)\rangle^2=\langle \hat S_{\theta}^2(\tau)\rangle$,} normalized by its bare vacuum counterpart.  Both generalized quadratures, corresponding to $\theta=0$ and $\theta=\pi/2$, respectively, are shown in dependence on the time delay $\tau$.
    The cat state is composed of a superposition of two broadband coherent states, with opposite amplitudes having Gaussian frequency distributions of  central frequency $\Omega_\mathrm{THz}=0.26 \omega_0$ and width $\sigma_\mathrm{THz}=0.13 \omega_0$, 
    so that the mean photon number is $ \langle \text{cat}|\hat{N} |\text{cat}\rangle=$ 1. For detection, $\omega_\mathrm{cut}=2\omega_0$ is used.}
    \label{fig:S2}
\end{figure}
{Measuring at $\theta=\pi/2$ ($\phi=0$), we do not get a direct access to the background NIR vacuum signal $\langle \hat{S}_{B,\mathrm{cut}}^2\rangle$ for the self-referencing. However, its contribution
to the calculated signals is the same for both generalized quadratures. Therefore,
measurements for both $\theta=0$ and $\theta=\pi/2$, with and without the frequency cut, provide sufficient information
to calculate
$\langle \hat{S}_{B,\mathrm{cut}}^2\rangle=\langle \hat{S}_{B}^2\rangle\langle \hat{S}_{\mathrm{hom,cut}}^2\rangle/\langle \hat{S}_{\mathrm{hom}}^2\rangle$
 required for the self-referencing, up to generally minor corrections due to a change in the spectral shape of $\mathcal{G}_{\phi=-\pi/2}(\Omega)$ caused by the cut.}

\section{Results}
Let us first illustrate
how the proposed scheme for sampling of quantum fields operates in the case when we apply it to the bare THz vacuum, using typical experimental parameters.
We assume that the probe is Gaussian,
has central frequency $\omega_0$ and spectral width $\sigma$. The LO field
has central frequency $2\omega_0$ and spectral width $\sqrt{2}\sigma$.
Ignoring dispersion and assuming a flat response for the detector crystal, i.e., $n(\omega_2)=n$ and $\eta(\omega_2)=\eta$ as well as no spectral cuts for the detected photons, meaning $\omega_\mathrm{cut}=\infty$ in Equation \eqref{Eq:G},
gives
$\mathcal{G}_{\pm}(\Omega)=\eta \exp{\left(-\Omega ^2/8 \sigma ^2\right)}/2\sqrt{2\pi } \sigma.$
The TFISH contribution to the variance is maximized for
$\phi=-\pi/2$, with $\mathcal{G}_{\phi}(\Omega)=2\mathcal{G}_{+}(\Omega)$ here. We get then
\begin{equation}\label{Eq:variance_vac_result}
\langle \hat{S}^2\rangle_{\rm vac}
=
\frac{9 d^2 \eta ^2 \eta _2 (1-R) R^2 \sigma ^3 (\chi^{(3)})^2 \omega _0^3 \hbar ^2 }{8 \pi ^{3/2}  c^4 n^4 \epsilon _0^2  F^2}
 N_P^3,
\end{equation}
where
$
 N_P=  A_P^2 n C'\int_0^\infty \frac{f^2(\omega) d\omega}{\hbar \omega}\approx  \frac{  A_P^2 n C'}{2\sqrt{\pi} \hbar \sigma \omega_0 }
$
is the average number of  photons in the probe.
%
Under the same conditions, we obtain
\begin{equation}
\langle \hat{S}_B^2\rangle\approx A^2 C'  \frac{n \eta^2}{2\sqrt{ \pi}\hbar \sigma \omega_0}
= \eta^2\eta_2 (1-R) N_P.
\end{equation}

\textbf{Figure \ref{fig:S2}}a shows $\langle \hat{S}^2\rangle_{\rm vac}$, $\langle \hat{S}_B^2\rangle$ and the total variance $\langle \hat{S}_\mathrm{hom}^2\rangle$ as functions of the number of photons in the probe pulse, contrasted with an estimation of
the back-action efficiency
of the $\chi^{(3)}$ process in the perturbative regime
(see Supporting Information for details).
We choose  the parameters considering a probe laser with 1.55~$\mu$m central wavelength and 6~fs pulse duration and a thin 12~$\mu$m slab of diamond for the $\chi^{(3)}$ crystal.
The variance of the TFISH part surpasses the variance due to the background vacuum at around $N_P=3.7\times 10^{11}$ photons per pulse. However, the back-action becomes significant around the same photon number, thus invalidating the perturbative approach.
To prevent this, $N_P$ can be chosen at a lower value, implementing a weak measurement.  To achieve the required signal-to-noise ratio, the measurement results can be then averaged over a sufficient number of probe pulses, as in \cite{Riek2015,Riek2017,Chelmus2019}.

As an illustration of the scheme going beyond sampling of the THz quantum vacuum, \textbf{Figure \ref{fig:S2}}b shows the variance signals for a broadband cat state, as a function of the time delay of the probe. The utilized cat state is defined as $|\text{cat}\rangle \propto   |\{\alpha_\Omega\}\rangle + |\{-\alpha_\Omega\}\rangle  $, where $|\{\alpha_\Omega\}\rangle$ represents a continuous multimode coherent state with the spectral amplitudes
$\alpha_\Omega$ corresponding to a classical
few-cycle THz pulse \cite[p.~85]{Vogel_book}. To access both quadratures, we performed the frequency cut on the spectral content of the photons collected by the photodetectors, determining the gating function, Equation \eqref{Eq:G}, by choosing $\omega_\mathrm{cut}=2\omega_0$. Further details on the calculation are given in Supporting Information.

\section{Conclusion}
We have developed a new scheme for sampling of quantum fields in the time domain and illustrated it by calculating the electric-field variance and its dynamics for the quantum vacuum and a pulsed broadband cat state, respectively. The scheme is feasible for typical experimental parameters and has a number of intrinsic advantages, such as automatic subtraction of the contaminating shot-noise by the shot-by-shot self-referencing. High sensitivity in the 5-15 THz provides the ability to study low-energy quantum dynamics in condensed matter, while 
frequency filtering of homodyne signal gives access to both generalized electric-field quadratures, en route toward subcycle quantum tomography. Finally, lifting the reliance on high-precision polarization optics holds promise for future imaging and microscopy applications involving THz and MIR quantum fields.

%
%
%
%
%
%

\medskip
\textbf{Supporting Information} \par 
Supporting Information is available from the Wiley Online Library or from the author.

\medskip
\textbf{Acknowledgements} \par 
This research was supported by the National   Research   Foundation of   Korea  (NRF)  grant  funded  by  the  Korea  government (MSIT)(2020R1A2C1008500).
A.S.M was also supported by the  Mercator Fellowship of the Deutsche Forschungsgemeinschaft (DFG) - Project No.   425217212 - SFB 1432 and Baden-W\"{u}rttemberg  Stiftung  via the   Elite  Programme   for Postdocs.
S.G. acknowledges support by the Institute for Basic Science in Korea (IBS-R024-D1). S.V., M.S. and D.V.S. acknowledge funding by  the Natural Sciences and Engineering
Research Council of Canada (NSERC) via the Canada Research Chair (CRC) of D.V.S. and the Fonds de Recherche
du Qu\'{e}bec –Nature et Technologies (FRQNT) via Institut Transdisciplinaire d{'}Information Quantique (INTRIQ). Further, this project has received funding from the European Union’s Horizon Europe research and innovation programme under grant agreement No. 101070700.

\medskip

\end{document}